%% file: main.tex
\documentclass[journal=psabav,manuscript=article]{achemso}

\usepackage[T1]{fontenc} % Use modern font encodings
\input{packages}

\input{macros}

\author{Tobias~Eklund}
\email{teklund@uni-mainz.de}
\affiliation{Institute of Physics, Johannes Gutenberg University Mainz, Germany}
\alsoaffiliation{European XFEL GmbH, Schenefeld, Germany}
\alsoaffiliation{Physics at Interfaces, Max Planck Institute for Polymer Research, Mainz, Germany}
\author{Christina~M.~Tonauer}
\affiliation{Institute of Physical Chemistry, University of Innsbruck, Austria}
\alsoaffiliation{Deutsches Elektronen-Synchrotron DESY, Hamburg, Germany}
\author{Felix~Lehmkühler}
\affiliation{Deutsches Elektronen-Synchrotron DESY, Hamburg, Germany}
\author{Katrin~Amann-Winkel}
\affiliation{Institute of Physics, Johannes Gutenberg University Mainz, Germany}
\alsoaffiliation{Physics at Interfaces, Max Planck Institute for Polymer Research, Mainz, Germany}

\begin{document}

% \begin{tocentry}

% Some journals require a graphical entry for the Table of Contents.
% This should be laid out ``print ready'' so that the sizing of the

% Inside the \texttt{tocentry} environment, the font used is Helvetica
% 8\,pt, as required by \emph{Journal of the American Chemical

% The surrounding frame is 9\,cm by 3.5\,cm, which is the maximum
% permitted for  \emph{Journal of the American Chemical Society}
% graphical table of content entries. The box will not resize if the
% content is too big: instead it will overflow the edge of the box.

% This box and the associated title will always be printed on a
% separate page at the end of the document.

% \end{tocentry}

\begin{abstract}
\input{section_text_files/abstract}
\end{abstract}


\input{section_structure}

\begin{acknowledgement}
\input{section_text_files/acknowledgement}
\end{acknowledgement}

% \begin{suppinfo}

% A listing of the contents of each file supplied as Supporting Information
% should be included. For instructions on what should be included in the
% Supporting Information as well as how to prepare this material for
% publications, refer to the journal's Instructions for Authors.

% The following files are available free of charge.
% \begin{itemize}
%   \item Filename: brief description
%   \item Filename: brief description
% \end{itemize}

% \end{suppinfo}

%%%%

% \section{Funding}

% \section{Conflicts Of Interest}

\newpage
\bibliography{contin}

\end{document}

%% file: packages.tex
\usepackage{amsmath}
\usepackage{hyperref}

%% file: macros.tex
\newcommand{\isf}{f} %intermediate scattering function
\newcommand{\dd}{\,\mathrm{d}} %differential element
\newcommand{\e}{\mathrm{e}} %eulers const.
\newcommand{\Ddist}{\Phi} %diffusive comp. dist symbol
\newcommand{\vdist}{\Psi} %Ballistic comp. dist symbol
\newcommand{\matr}[1]{\boldsymbol{\mathbf{#1}}}  %matrix symbols
\newcommand{\lognorm}{\mathcal{N_L}} %pdf of log normal dist.
\newcommand{\pollard}{\mathcal{P}} %pdf of Pollard's dist.

\title{Global $q$-dependent inverse transforms of intensity autocorrelation data}

%% file: section_text_files/abstract.tex
We present a new analysis approach for 
intensity autocorrelation data, as measured with dynamic light scattering and X-ray photon correlation spectroscopy. Our analysis generalizes the established CONTIN and MULTIQ methods by direct nonlinear modeling of the $g_2$ function, enabling decomposition of complex dynamics without a priori knowledge of experimental scaling factors. We describe the mathematical formulation, implementation details, and strategies for solution, as well as demonstrate decompositions of soft matter dynamics data into distributions of diffusion rates/velocities. The open-source MATLAB implementation, including example data, is publicly available for adoption and further development.

%% file: section_structure.tex
\section{Introduction}
\label{sec:intro}
\input{section_text_files/intro}

\section{Problem setup, array structure and solution}
\label{sec:setup}

\input{section_text_files/setup}

\section{Choosing the regularization parameter \texorpdfstring{$\lambda$}{\em{λ}} and generalized degrees of freedom}
\label{sec:reg_and_dof}

\input{section_text_files/reg_and_dof}

\section{Example applications}
\label{sec:exp_examples}

\input{section_text_files/exp_examples}

\section{Special simplifying cases}
\label{sec:simplified_methods}

\input{section_text_files/special_cases}

\section{Conclusions}
\input{section_text_files/conclusions}

\section{Data Availability}
\input{section_text_files/data_availability}

%% file: section_text_files/intro.tex
Dynamic light scattering (DLS)\cite{berne_dls} and X-ray photon correlation spectroscopy (XPCS)\cite{lehmkuehler_2021,sandy_2018,madsen_2018,shpyrko_2014,sinha_2014,perakis_2020,grubel_2004} are powerful techniques for probing the dynamics of soft and hard matter systems. 
Both methods exploit the temporal fluctuations of scattered light intensity to access information about relaxation processes and microscopic motion.
In DLS, visible or near-infrared laser light is scattered by suspended particles, while XPCS extends the technique to the X-ray regime, enabling studies of optically opaque materials and smaller length scales. Recent developments on  diffraction-limited storage rings and X-ray free electron lasers enable exploration of timescales ranging from hours down to just a few femtoseconds.\cite{lehmkuehler_2021,madsen_2018}
The key experimental observable is the intensity autocorrelation, $g_2(q,\tau)$, measured as a function of the scattering vector $q$ and correlation lag time $\tau$ (in terms of experimental geometry, the length of the scattering vector is $q = 4\pi\sin(\theta/2)/\lambda$, where $\theta$ is the scattering angle and $\lambda$ the wavelength of the incident radiation). Analysis of $g_2$ can give access to the intermediate scattering function, $f(q,\tau)$, a complete account of density correlations across (reciprocal) space and time. As such, the results from DLS and XPCS provide a window into various dynamic processes, including diffusion, flow, and structural relaxation.

Characterizing dynamics from temporal correlation data
may require inversion of integral transforms, as the intermediate scattering often takes the form of
superimposed decays.
This could be the case, for example, when disperse scatterers undergo Brownian diffusion at different diffusion rates. The intermediate scattering function for a uniform population is\cite{berne_dls}
\begin{equation}
    f(q,\tau) =  \e^{-q^2 D \tau}.
\end{equation}
Assuming that these field auto-correlations combine additively, a disperse sample gives dynamic scattering of the form
\begin{equation}
    \label{eq:diff_f}
    f(q,\tau)=\int\Ddist(D) \e^{-q^2 D \tau}\dd D.
\end{equation}
Here, $\Ddist(D)$ represents the density of particles with diffusion coefficient $D$. The challenge for data analysis is, then, to recover the distribution of diffusive modes from an experimental estimate of $f$. The standard techniques are empirical curve-fitting (e.g., fitting stretched exponentials to $f$),\cite{lehmkuehler_2021} estimation of distribution cumulants,\cite{koppel_1972} or computing the density function $\Ddist$ by inverting the above integral transform\cite{provencher_1982a}.

The CONTIN Fortran program has been used to find the inverse transform since its conception in the late 1970s~\cite{provencher_1979}. This inversion problem is ill-posed in the sense that many qualitatively distinct solutions $\Ddist$ may satisfy equation \eqref{eq:diff_f} to within acceptable error. Naive approaches towards optimization will generally favor highly complex solutions that fit closely to not only sample dynamics, but also to any experimental noise in $f$. \citeauthor{provencher_1979}\cite{provencher_1979} applied ridge regression (a young technique at the time) to select the least complex solution consistent with the data, thereby taming the inverse transform problem. CONTIN is, to this day, the standard software library used for this type of analysis.

The first version of CONTIN solved equation~\eqref{eq:diff_f} without consideration of the $q$-dependence (i.e., solving $f(\tau)=\int\Ddist(\Gamma)\exp{(-\Gamma\tau)}\dd\Gamma$ independently for discrete $q$ values). A later addition, called MULTIQ~\cite{provencher_1996}, introduced the idea of a global $q$-dependent inverse transform. The original application was to resolve a diffusive and a relaxational ($q$-independent) component,
\begin{equation}
    \label{eq:provencher_2comp}
    f(q,\tau) = \underbrace{\int\Ddist(D)\e^{-q^2 D\tau}\dd D}_{\text{Diffusive component}} + \underbrace{\int \Theta(\Gamma)\e^{-\Gamma\tau}\dd \Gamma.}_{\text{Relaxational component}}
\end{equation}
By proposing a stronger hypothesis, including the $q$-dependence, and performing a global analysis ($q$\nobreakdash-$\tau$~surface fit), distributions of decays that appear monomodal at any given scattering vector could be resolved into distinct distributions of diffusive vs. relaxational modes.

There has been a recent resurgence of interest in the inverse transform method for data analysis, driven in part by the advent of modern high brightness X-ray sources and the consequent boost for coherence-based techniques such as XPCS. \citeauthor{andrews_2018a} developed CONTIN further for X-ray applications, and introduced a diffusive-ballistic model for MULTIQ~\cite{andrews_2018a}. This brought the inverse transform method to attention in scientific communities that have traditionally relied on the curve-fitting method. \citeauthor{marino_2007_mlfe}, and later, \citeauthor{lienard_2022} re-built some of CONTIN's functionality in MATLAB.\cite{marino_2007_mlfe,lienard_2022,lienard_2022_mlfe} Re-implementation in a modern, interpreted language is a well-intentioned move which increases the ease-of-use and flexibility for scientists who may find CONTIN's Fortran code base difficult to work with. This, in our opinion, is the right way forward, but as far as we know, no modernized, freely available remake implements the essential MULTIQ functionality.

The intermediate scattering function can be measured by exploiting the so-called Siegert relation. Siegert's theorem states that the intensity autocorrelation,
\begin{equation}
    g_2(q,\tau)=\frac{\langle I(q,t)I(q,t+\tau)\rangle_t}{\langle I(q,t) \rangle_t^2},
\end{equation}
of a speckled interference pattern is related to the intermediate scattering function as\cite{voigt_1994}
\begin{equation}
    \label{eq:siegert}
    g_2(\vec q,\tau) = 1+\beta \left|\isf(\vec q,\tau)\right|^2
\end{equation}
(some example $g2$ curves are shown in figure \ref{fig:g2_bl_contr}). The instrumental factor $\beta$, the speckle contrast, is related to the coherence properties of the illuminating beam and the size of the detector. Since the intensity can be measured at various scattering angles simultaneously with a 2D detector (CCD, photodiode array, hybrid pixel detectors, etc.), an estimate of $f$ is, in principle, easily obtained by rearranging Siegert's statement into
\begin{equation}
    \label{eq:siegert_f}
    \left|\isf\right| = \sqrt{\frac{g_2-1}{\beta}}.
\end{equation}
This can then be given as input data to CONTIN (or similar solvers). 

In practice, measurements are often afflicted with parasitic scattering and various optical issues that cause deviations from equation \eqref{eq:siegert}. Such deficiencies can be difficult to mask out by hand, but in most cases, the systematic errors they cause can be modeled as a time-constant, per $q$, baseline and $q$-dependent contrast,\cite{provencher_1978}
\begin{equation}
    g_2(q,\tau)=1+b(q)+\beta(q) \left|\isf(q,\tau)\right|^2.
\end{equation}
To further compound the issue, correlation times in glasses and amorphous solids may be several minutes.\cite{ladd_parada_2022} Collecting sufficient statistics for such long delay times is not always practical during time-limited synchrotron measurements, resulting in a truncated tail as in figure \ref{fig:g2_bl_contr}. This rules out establishing independent estimates of the free baseline and contrast parameters.
Whenever $b$ and $\beta$ are known, $|f|=\sqrt{(g_2-1-b)/\beta}$ can stand in for equation \eqref{eq:siegert_f}. When they are unknown, they need to be evaluated in a best-fit manner, together with a hypothesis for $f$. This leads to a nonlinear model of the form
\begin{equation}
    \label{eq:general_g2}
    g_2(q,\tau) = 1+ b(q)+\beta(q)\left( \sum_l\int\phi_l(s) \e^{- q^{\zeta_l} (s\tau)^{\eta_l}}\dd s\right)^2.
\end{equation}
This expression covers the diffusive-relaxational case in equation \eqref{eq:provencher_2comp}, and a ballistic component (with mean squared displacement proportional to $\tau^2$) is achieved by setting $\zeta_l=\eta_l=2$.

\input{figures_tex/fig_g2_bl_contr}

\newpage

Various ways of dealing with equation \eqref{eq:general_g2} have been proposed. Some earlier work applies the CONTIN method directly to $g_2$~\cite{andrews_2018a,lienard_2022}. This amounts to a linearizing approximation where squaring the transform kernel(s) stands in for squaring all of $f$,
\begin{equation}
    \left( \sum_l\int\phi_l \e^{- q^{\zeta_l} (s\tau)^{\eta_l}} \dd s\right)^2 \approx \sum_l\int\phi_l(s) \e^{- 2 q^{\zeta_l} (s\tau)^{\eta_l}}\dd s.
\end{equation}
This relation is exact for the case of a single, sharply monomodal component, but will generally lead to an over-interpretation in terms of size dispersity and/or dynamical heterogeneity (cross-terms from squaring are interpreted as sample  dynamics). Another option is to solve the $g_2$ problem directly; inverting equation \eqref{eq:general_g2} to recover the functions $\phi_l$. This requires general nonlinear programming (NLP) techniques, and is not supported by CONTIN. Previous work on this nonlinear problem, for example that of \citeauthor{weese_1993},\cite{weese_1993} does not implement the MULTIQ functionality.

We here present new software designed for solving equation \eqref{eq:general_g2} with a regularized nonlinear fitting method. This code is derived from the previous efforts by \citeauthor{marino_2007_mlfe} and \citeauthor{lienard_2022},\cite{marino_2007_mlfe,lienard_2022_mlfe,lienard_2022} but supplemented with the following features:
\begin{itemize}
    \item Nonlinear modeling of the $g_2$ intensity autocorrelation, according to equation \eqref{eq:general_g2}.
    \item Global analysis with $q$-dependent transform kernels, cf. CONTIN MULTIQ~\cite{provencher_1996}.
    \item Estimation of statistical degrees of freedom with Ye's generalized degrees of freedom~\cite{ye_1998}.
    \item Selecting an appropriate level of regularization with Provencher's 50\% rejection criterion~\cite{provencher_1982a}.
\end{itemize}
Like the previous work, ours is a MATLAB implementation, using solver tools from the Optimization Toolbox. In the next section, we discuss the program architecture. After that follows, in order: some details on degrees of freedom and regularization, demonstration of two soft matter applications , and, finally, two simplified methods for treating special cases.

%% file: figures_tex/fig_g2_bl_contr.tex
\begin{figure}
    \includegraphics{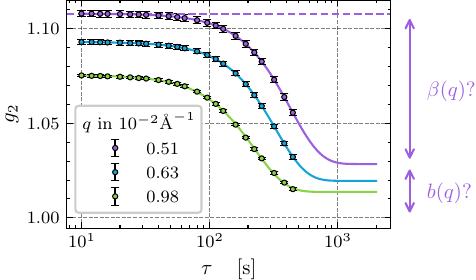}
    \caption{Example $g_2$ curves from a measurement on amorphous ice taken at 110 K temperature. In this example, the very long delay times ($>10^3$ seconds) are missing, giving a truncated tail, and a priori unknown values for baseline $b$ and contrast $\beta$.}
    \label{fig:g2_bl_contr}
\end{figure}

%% file: section_text_files/setup.tex
In this section, we describe the basic workings of our program, beginning with defining our notation and conventions. We refer to numeric arrays by upright bold symbols ($\matr{T}$). We use indexed italics for individual elements of arrays, e.g., ${g_2}_{ij}$ is the element from row $i$, column $j$ of the array $\matr{g_2}$. Repeated indices do \underline{not} represent contractions. A hat symbol is used to distinguish input data, $\matr{\hat{g}_2}$, from model values, $\matr{g_2}$.
The input correlation data (and corresponding model response) is arranged in a 2D array according to ${\hat{g}}_2(q_i,\tau_j)={{\hat{g}}_2{}}_{ij}$, i.e., with $q$ constant along rows, and $\tau$ constant along columns. We refer to the number of $q$ bins and delay times in the input data as $Q$ and $N$, respectively. The total number of dynamical components, the integral terms in equation \eqref{eq:general_g2}, is $L$. The arrays in our code are structured as described in this text. 

Initially, our setup follows the conventional approach~\cite{provencher_1982a, weese_1993, andrews_2018a,lienard_2022}. The integrals of equation \eqref{eq:general_g2} are discretized according to
\begin{equation}
    \label{eq:discritization}
    \sum_{l=1}^L\int \phi_l(s) \e^{- q^{\zeta_l}(s \tau)^{\eta_l}}\dd s \approx
	   \sum_{m=1}^{LM} {\Ddist}_{m}
	   \mathrm{e}^{- q^{\zeta_{l'}}(s_{m}\tau)^{\eta_{l'}}}w_{m}\quad 
       % \left(\text{with } l'=\left\lfloor\frac{M+m-1}{M}\right\rfloor\right),
       \left(\text{with } l'=\left\lceil\frac{m}{M}\right\rceil\right),
\end{equation}
giving $M$ fit parameters $\Ddist_{m}$ per component to solve for (total $LM$). Each coefficient $\Ddist_{m}$ is an approximation of the density function evaluated at the corresponding point; $\Ddist_{m} \approx \phi_{l'}(s_{m})$. The model intermediate scattering function is computed as matrix multiplication with a design array $\matr{T}$;
\begin{equation}
    f_{ij}= \sum_{m=1}^{ML} T_{imj}\Ddist_{m} =
       \begin{pmatrix}
        \e^{- q_i^{\zeta_1}(s_1 \tau_j)^{\eta_1}}w_{1} &
        \ldots &
        \e^{- q_i^{\zeta_L}(s_{LM} \tau_j)^{\eta_L}}w_{LM} &
        \end{pmatrix}
    \begin{pmatrix}
        \Phi_{1} \\
        \vdots      \\
        \Phi_{LM}
    \end{pmatrix}.
\end{equation}
The array $\matr{T}$ (of shape $Q\times ML \times N$) is stored in memory and reused for every call to compute $\matr{f}$. For simplicity, we use the rectangular rule for the weights $w_{m}$, but other methods can be considered for increased precision. Some choices, e.g., Gauss-Legendre quadrature, may restrict the choice of evaluation points $s_{m}$ in ways that are unpractical. In our experience, convergence is more reliable and quicker if the evaluation points are logarithmically spaced (which we set as the default behavior). Finally, the model for $g_2$ is constructed from the $\matr{f}$ array as
\begin{equation}
    {g_2}_{ij} = 1 + b_i + \beta_i f_{ij}^2.
\end{equation}
(this introduces $2Q$ more fitting parameters $b_i$, $\beta_i$). Here, we have departed from the linear approach (CONTIN and derivatives). We make no attempt at turning this into a linear model; $g_2$ is computed  element-wise according to Siegert's formula with the baseline and contrast corrections, including all cross-terms that arise in the squaring of $f_{ij}$. 

To condition the $\matr{T}$ array, we normalize $q$ and $\tau$. Our standard choice is to give delay and momentum transfer in units of the largest $\tau$ and $q$ in the dataset,
\begin{equation}
    \label{eq:data_norm}
    q_i=\frac{q'_i}{q'_\text{max}}\qquad \tau_j =\frac{\tau'_j}{\tau'_\text{max}}
\end{equation}
(with $q'_i$, $\tau'_i$ as the original data coordinates). With this choice, $s=1$ gives a decaying mode that reaches the $1/\mathrm{e}$ level at the highest (usually fastest) $q$ and longest delay time. So, $s=1$ is quite slow with respect to the available data. The quantities $s$, and $\phi_l(s)$ can be converted into a rate in the original units via the variable change
\begin{equation}
    \begin{cases}
        D = \frac{s}{{q'_\text{max}}^{\zeta_l/\eta_l} {\tau'_\text{max}}} \\
        \phi_l(s)\dd s = \phi_l\left({q'_\text{max}}^{\zeta_l/\eta_l}{\tau'_\text{max}}D\right)\;{q'_\text{max}}^{\zeta_l/\eta_l}{\tau'_\text{max}} \dd D.
        %\phi_l(s)\dd s =\phi_l(q_\text{max}^\xi\tau_\text{max}^\eta D^\eta) q_\text{max}^\xi\tau_\text{max}^\eta \eta D^{\eta-1}\dd D.
    \end{cases}
\end{equation}
This gives a distributions of diffusion rates (length squared over time) in the diffusive case, and velocities (length over time) for the ballistic case.

All the free parameters are held in a solution vector of length $2Q+LM$,
\begin{equation}
    \matr{x}=
    \begin{pmatrix}
    b_1 & \ldots & b_Q & \beta_1 & \ldots & \beta_Q & \Phi_{1} & \ldots & \Phi_{M} & \Phi_{M+1} & \ldots & \Phi_{LM}
    \end{pmatrix}^\mathrm{T}.
\end{equation}
These need to be constrained in order for the solver to converge on a reasonable solution. In experiments, static samples (e.g., Aerogel) are typically used to determine the maximum allowed contrast $\beta_i$. The linear equality  constraint
\begin{equation}
    \label{eq:isf_norm_constraint}
    \sum_m\Phi_{m}w_{m}=1
\end{equation}
% (eq. to $\int f()
enforces a normalized intermediate scattering function, and reduces the size of the solution space. We usually also set Neumann boundary conditions ($\Phi_{(l-1)M}=\Phi_{lM}=0$ for $l=1\ldots L$), such that long term excess in correlation is associated with the baselines $b_i$ rather than with slowly decaying modes. This equality constraint is only appropriate if the sample dynamics are expected to go to zero at the integral limits. Finally, it is useful to constrain the baseline parameters $b_i$ to stay within some reasonable bounds (usually $b_i\in (0,0.05)$ or similar).

The objective function to be minimized is
\begin{equation}
    \label{eq:objective_function}
    O(\matr{x}) = 
    \chi^2+\lambda R(\matr{x}).
\end{equation}
Here $\chi^2$ is the weighted sum of squared residuals over all $q$ and $\tau$ in the input data,%\sepfootnote{setup:uncertainties}
\begin{equation}
    \chi^2=\sum_{i=1}^Q\sum_{j=1}^N\left(
        \frac{{{\hat{g}}_2{}}_{ij}-{g_2}_{ij}}{\hat{\sigma}_{ij}}
    \right)^2.
\end{equation}
The weights are supplied by the user as a $Q\times N$ array. When analyzing XPCS data, we compute the autocorrelation for every detector pixel, then take the mean over pixels at roughly equivalent points in momentum space as our observations $\matr{\hat{g}_2}$ and the standard error of that mean as our uncertainty $\matr{\hat{\sigma}}$. The second term, $\lambda R$, is the regularization mechanism. Typically, $R$ is chosen such that it is large when the solution has many maxima, or exhibits other undesirable traits that may appear due to overfitting. The value chosen for the Lagrange multiplier $\lambda$ decides to what degree the side-constraint $R=0$ is violated, striking a balance between a closer fit  and a less complex solution. For more details about the role of the regularizer, see Provencher's original papers and \citeauthor{andrews_2018a}\cite{provencher_1979,provencher_1982a,andrews_2018a} An appropriate value for $\lambda$ is not a priori known, and has to be deduced from fit statistics, a process discussed in the next section.

For regularization, we use the popular second derivative norm,
\begin{equation}
    \label{eq:reg_multicomp}
    R=\sum_l\int \left(\frac{\dd^2 \phi_l(s)}{\dd s^2}\right)^2\dd s.
    % \quad\text{where }s = u^{\eta_l}.
\end{equation}
The accepted generalization to combinations of modes of different types (diffusive, ballistic, etc.) is to add the respective norms together as per the above equation~\cite{andrews_2018b}. The normalization from equation~\eqref{eq:data_norm} ensures that the relative weights of the regularizer terms are independent of the data units, but the size of each term still depends on how $s$ was defined to begin with. We have chosen to raise the integration variable to the same power as the time coordinate in our transform kernels, with decaying modes given by $\exp{(-q^{\zeta}(s\tau)^{\eta})}$, see equation \eqref{eq:general_g2}. This puts the regularizer terms on a sort of equal footing with respect to time. Another choice could have been $\exp{(-s q^{\zeta}\tau^{\eta})}$, but we found that this tends to punish components with larger $\eta$ less severely for exhibiting multiple maxima, resulting in solutions with a very jagged ballistic component, but an overly smooth diffusive component. With practically endless possibilities, we must concede that the choices reflected in equations \eqref{eq:general_g2} and \eqref{eq:reg_multicomp} are largely arbitrary. For applications with various $q$-powers, this specific setup could prove detrimental, and users may want to modify the kernel or regularizer definition.
\newpage

To carry out differentiation as a forward finite difference, we use a sparse, block tridiagonal matrix,
\begin{equation}
    \matr{\Phi''}=
    \underbrace{\begin{pmatrix}
        \begin{matrix}
            A_1 & B_1 & C_1 & 0   & 0 & 0 & 0 & 0\\
            0 &  \ddots&\ddots&  \ddots  & 0 & 0 & 0 & 0\\
            0 &   0 & A_M &B_M &   C_M & 0 & 0 & 0\\ 
           0 & 0 & 0 & 0 & 0 & 0 & 0 & 0\\
            0 & 0 & 0 & 0 & 0 & 0 & 0 & 0\\
            0 & 0 & 0 & 0 & 0 & A_{M+1} & B_{M+1} & C_{M+1}
        \end{matrix} & \ldots \\
        \vdots & \ddots \\
    \end{pmatrix}}_{\matr{\Delta}}
    \begin{pmatrix}
        \Phi_1 \\
        \vdots\\
        \Phi_{LM}
    \end{pmatrix}.
\end{equation}
Since a twice applied difference operator reduces the number of data points by two, we added two rows of zero-padding for each dynamical component. To account for possible unequal step sizes, the diagonals are given by
\begin{equation}
    \begin{cases}
    A_k=\frac{1}{ (s_{k+1}-s_k)(s_{k+2}-s_k)},  \\
    B_k=-\left(\frac{1}{s_{k+2}-s_{k+1}}+\frac{1}{s_{k+1}-s_k}\right)\frac{2}{(s_{k+2}-s_k)}, \\
    C_k = \frac{2}{(s_{k+2}-s_{k+1})(s_{k+2}-s_k)}.
    \end{cases}
    % \begin{cases}
    % A_k=&\frac{1}{ (u_{k+1}-u_k)^2},  \\
    % B_k=-&\frac{1}{(u_{k+1}-u_k)}\left(\frac{1}{u_{k+2}-u_{k+1}}+\frac{1}{u_{k+1}-u_k}\right) \\
    % C_k = &\frac{1}{(u_{k+2}-u_{k+1})(u_{k+1}-u_k)}
    % \end{cases}
\end{equation}
This gives the regularizer as
\begin{equation}
    \label{eq:reg_matr}
    R = 
    \sum_{m=1}^{LM} w_m\left(\Phi_m'' \right)^2=\sum_{m=1}^{LM}\sum_{n=1}^{LM} w_m\left({\Delta}_{mn}\Phi_n \right)^2.
\end{equation}
We chose the forward difference because a central difference, while more precise overall, can assign very small values to rapidly oscillating solutions.

 With the above discretization and regularized least squares formulation, the variational problem of finding the best fit functions $\phi_l$ is now a ``straight-forward'' constrained numerical optimization problem. The objective function $\eqref{eq:objective_function}$ is up to cubic in the fitting parameters (there are terms like $\beta_i \Phi_j \Phi_k$), and we have some linear side  constraints (equation \eqref{eq:isf_norm_constraint} and the parameter bounds). This is solved with a sequential quadratic programming (SQP) algorithm. Given a starting point, the problem is locally modeled by a quadratic approximation. Moving to the solution of the quadratic subproblem brings us closer to a local minimum of the full problem.  Each iteration in the sequence uses an active-set strategy to handle the constraints.\cite{nocedal_2006}
 
The SQP solver will find a local minimum (or fail to converge), but we want to find the global minumum. For that, we employ the ``multistart'' technique: a number of initial guesses are generated, and the optimization is run for each one in parallel. The solution with the lowest value for the objective function (and a positive exit flag from the solver) is picked as the best solution. For starting points, we use smooth, compactly supported bump functions of the type
\begin{equation}
    % \left\{
    \begin{cases}
        &\Ddist_{m} = A\e^{h^2/((m-c)^2-(h+1)^2)},\quad c-h\leq m \leq c+h\\
        &\Ddist_{m} = 0 \quad \text{elsewhere}
    \end{cases}
    %\right.
\end{equation}
where $c$ is the bump center and $h$ is the half-width of the supporting region.
The coefficient $A$ is fixed by the usual normalizing condition \eqref{eq:isf_norm_constraint}. We have found that starting from four evenly spaced, non-overlapping bumps usually leads to at least one sensible solution. If there is more than one dynamical component, we consider the possible pairs of bumps per component, and also include start points where all $\Ddist_{m}$ are set to zero for one component. This gives a total of $5^L-1$ different starting points (for $L=2$, we get 24 parallel processes). The initial values for contrast and baselines are chosen as the midpoint between the respective parameter bounds. It is very likely that this procedure can be improved to generate fewer, but better, guesses, which would reduce the overall computational requirements.

%% file: section_text_files/reg_and_dof.tex
We use the same hypothesis test as the original CONTIN to select the weight of the regularization term~\cite{provencher_1979,provencher_1982a,provencher_1982b}. Denoting the residual sum of squares in equation \eqref{eq:objective_function}, after optimization with a specific value for $\lambda$, by $\chi^2_\lambda$, the selection criterion is
\begin{equation}
    \label{eq:f-test}
    P\left(
        \frac{\chi^2_\lambda-\chi^2_0}{\chi^2_0}
        \frac{NQ-p_0}{p_0 } ; NQ-p_0,p_0\right)
    \approx 0.5,
\end{equation}
where $P$ is the CDF of Fisher's $F$-distribution, and $p_0$ are the regression effective degrees of freedom measured at small $\lambda$ (small enough to guarantee an over-fitted solution).  This quantity rises rapidly from nearly zero to one when the fractional increase in $\chi^2$ is no longer well explained by random errors. At $P>0.5$, the risk is high that the regularizer is imparting significant deviations from the data. Since $P$ is a monotonously increasing function of $\lambda$, we can easily find the wanted 50\% rejection level with a bisection search.

Figure \ref{fig:reg_and_dof} shows some example results with different values for the regularization parameter ($\lambda$). The sample is a dilute suspension of colloidal particles, exhibiting diffusive dynamics~\cite{frenzel_2021}. The $g_2$ model responses are nearly identical for moderate levels of regularization  (blue, purple and green $g_2$ curves all fall on top of each other), even though the interpretation in terms of diffusive modes transitions from a bimodal  to a monomodal distribution around $\lambda=10^{15}$ (purple vs. green curves). In this example, $\lambda=10^{15}$ (green) is close to the optimal 50\% rejection level. The orange curves ($\lambda = 10^{17}$) show an example of over-smoothing. This solution does not capture the data accurately according to the $F$-test ($P\approx 1$), and we can also see that the residual sum of squares (bottom right panel) and $g_2$ curves start to deviate significantly.

It is important to note that the regularizer induces a bias towards faster dynamics (the orange $g_2$ curves in figure \ref{fig:reg_and_dof} all decay faster than the data). This shows that the second derivative norm is not an ideal regularizer for defeating over-fitting to random noise (i.e., the deviations from the data are not random at all). However, the $F$-test helps with finding a minimally complicated solution with tolerable amounts of this fast-bias. A third derivative norm regularizer does not exhibit the same bias, but in our testing, restricting the third derivative  could never trigger the 50\% rejection level of the $F$-test. Such a regularizer can be minimized as far as the optimization constraints allow without disturbing the fit significantly. Regulating the third derivative with some other selection criterion may be a more natural choice, but this topic requires more investigation. In our implementation we keep the well used and established second derivative norm regularizer and $F$-test based selection criterion.

\input{figures_tex/fig_reg_and_dof}

To pick $\lambda$ with fit statistics, we need to know the statistical degrees of freedom. This is easy to find for a linear model (ordinary least squares). Each linear constraint reduces the degrees of freedom by one, leading to the famous formula
\begin{equation}
    \nu = NQ - p
\end{equation}
for the residual degrees of freedom with $NQ$ observations and $p$ fitting parameters. But, we are now in the nonlinear realm where fit parameters are not one-to-one with linear constraints (methods used for ridge regression do not apply in an obvious way either). Ye~\citeyear{ye_1998} has proposed a ``generalized degrees of freedom'', a measure of the flexibility of a  modeling procedure, for just this type of tricky case. We refer to the original  paper for mathematical detail, and here just give Ye's algorithm for computing the cost in degrees of freedom for our nonlinear modeling procedure.

\vspace{1em}
{\em Ye's algorithm}
\begin{enumerate}
    \label{}
    \item Generate $K$ sets of randomly perturbed data ${{\hat{g}}_2{}}_{ij}+\delta_{ijk}$, $k=1,\ldots,K$ 
    \item Solve the optimization problem for the perturbed data (resulting in $K$ model responses ${g_2}_{ijk}$)
    %\item Repeat from 1 as needed  to collect statistics (can be parallelized)
    \item Find the regression slops $p_{ij}$ from  ${g_2}_{ijk} = a_{ij} + p_{ij} \delta_{ijk}$
    \item The cost in DOF is $\sum_{i,j} p_{ij}$, giving 
    % \begin{equation}
        $\nu = NQ-\sum_{i,j}p_{ij}$
    % \end{equation}
\end{enumerate}
So, the regression degrees of freedom are measured by testing how the modeling procedure responds to variations in the data. In step 1, we use random variables with standard deviation $0.6\cdot\hat{\sigma}_{ij}$. Step 2 refers to the global solution, including the multistart procedure. This computation can be extremely costly, depending on the number of perturbations $K$ and start-points. Luckily, the computations are independent and can easily be parallelized.

We computed Ye's GDF (denoted by $p$) for a series of $\lambda$ values. The results can be seen in the bottom left panel of figure \ref{fig:reg_and_dof}. There are four interesting observations to be made here. First, the regression degrees of freedom is much smaller than the number of fitting parameters. In this case $p_0\approx 24$, for a 322 parameter model. This is similar to earlier estimates for ridge regression. Also, $\nu\ll NQ$ is an important requisite for the $F$-test method~\cite{provencher_1979}. Second, DOF only depends  weakly on $\lambda$ for moderate levels of regularization. This means that a constant or linear model for $p$ could be used to avoid costly recomputations of $p_\lambda$. Third, $p$ approaches zero for extremely smooth solutions. This is expected, as the heavy regularization should make the modeling procedure insensitive to variations in the data. Finally, since $p$ is estimated by a Monte Carlo method, there is some uncertainty and noise on the result. Increasing the number of data perturbations ($K$ in Ye's algorithm) reduces this noise at the cost of increased computation time. This example is computed with $K=93$, but we find that $K=50$ is typically enough get a consistent estimate of $\nu$. Overall, we find the results convincing, and use Ye's method to estimate $p_0$ in equation \eqref{eq:f-test}.

 Other methods for selecting $\lambda$ may of course be considered. We have tried the criterion $\chi^2/\nu\approx 1$, but found that this tends towards overly smooth solutions, which also exacerbates the regularizer bias discussed above. This is likely caused by unsuitable uncertainty estimates, which may include systematic errors (differences in $g_2$ within the $q$-bin) as  well as the expected random errors. The $F$-test is less sensitive to this issue~\cite{provencher_1979}. Since the DOF are costly to estimate, we advice against  methods that require recomputation of DOF at every $\lambda$. Provencher's $F$-test is better than the precise formula
 \begin{equation}
     P\left(
        \frac{\chi^2_\lambda-\chi^2_0}{\chi^2_0}
        \frac{NQ-p_0}{p_0-p_\lambda } ; NQ-p_0,p_0-p_\lambda\right)
 \end{equation}
 for this very reason. Some preliminary testing showed no significant difference between using the criterion \eqref{eq:f-test} and the classical $F$-test, so the (very long) time invested in recomputing DOF for every $\lambda$ had no real benefit. 

%% file: figures_tex/fig_reg_and_dof.tex
\begin{figure}
    \includegraphics{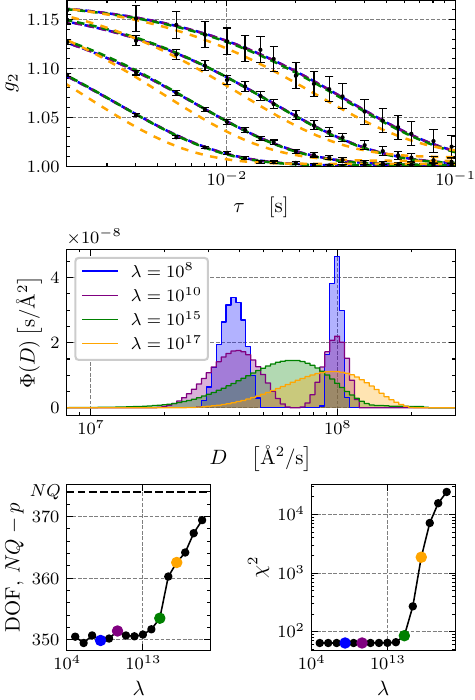}
    \caption{Example decompositions into diffusive modes, $\smash{f=\int\Ddist\e^{-Dq^2\tau}\;\dd D}$, with different regularizer weightings. The data is from an XPCS measurement of dilutely suspended colloidal particles. 
    \underline{Top}: Intensity autocorrelation data and lines of fit (dashed) from four select $q$ values (4.4, 6.5, 9.5, 15$\cdot10^{-4}$~Å\textsuperscript{-1}).
    \underline{Middle:} Solution distributions of diffusive modes. \underline{Bottom left:} Degrees of freedom estimated by Ye's algorithm. \underline{Bottom right:} Residual sum of squares.
    The blue and purple curves/points ($\lambda=10^8$, $10^{10}$) show results from lightly regularized optimization. The green curve ($\lambda=10^{15}$) is close to optimal according to Provencher's $F$-test, eq. \eqref{eq:f-test}. The orange curve ($\lambda=10^{17}$) is overly regularized. The purple, blue and green $g_2$ lines of fit are virtually identical at this scale, and all fall on top of each other.}
    \label{fig:reg_and_dof}
\end{figure}

%% file: section_text_files/exp_examples.tex
In this section, we show two example applications of our software. The first is an analysis of a system of colloidal particles (dilute suspension of silica-PNIPA, 9 wt\%). This data is from a previous study by \citeauthor{frenzel_2021}\cite{frenzel_2021} Our second example is a direct measurement on an amorphous solid, so-called amorphous solid water (ASW). For more information about ASW, see for example \citeauthor{burton_1935}, \citeauthor{mayer_1986} or \citeauthor{li_2021}\cite{burton_1935,mayer_1986,li_2021}  Both datasets are acquired with XPCS at the P10 beamline of PETRA III, DESY, Hamburg, Germany. These examples are included in our code repository~\cite{g2_decomposer_gh}.

\input{figures_tex/fig_PRE2021_2comp}

Figure \ref{fig:PRE2021_2comp} shows results for colloidal particles at two temperatures; 288 K and 313 K. The system undergoes a glass-liquid transition at 306 K, so there is an a priori expectation to find glassy (i.e. ballistic) dynamics at the low temperature and liquid-like (i.e. diffusive) dynamics at the higher temperature. In the first row of the figure, the intensity autocorrelation is plotted against a scaled time axis. Plotting against $q\cdot\tau$ makes the low temperature data for different $q$ values fall on top of each other, confirming our suspicion of relaxation rates proportional to $q$ (ballistic). Also note that the decays are distinctly Gaussian in shape close to $\tau=0$. Conversely, the higher temperature data (top right of figure \ref{fig:PRE2021_2comp}) is exponentially decaying with relaxation rates proportional to $q^2$ (diffusive), which is exposed when plotted against $q^2\cdot\tau$.

\newpage
As a sanity test for the fitting algorithm, we fit a mixed model of diffusive/ballistic dynamics given by
\begin{equation}
    \label{eq:ball_diff_g2}
    f(q,\tau)=
        \underbrace{
            \int\Ddist(D)\e^{-q^2D\tau}\dd D
            }_{\text{Diffusive component}} +
        \underbrace{
            \int\vdist(v)\e^{-\left(qv\tau\right)^2}\dd v.
            }_{\text{Ballistic component}}
\end{equation}
The  diffusive part should be mostly rejected at 288 K (i.e., $\int\vdist \dd v \approx 1$), while the ballistic part should be rejected at 313 K ($\int\Ddist \dd D \approx 1$). Rows two, three and four of figure \ref{fig:PRE2021_2comp} shows the results of our analysis. At 288~K, we find a smooth distribution of ballistic modes accounting for 76\% of the total dynamical content. The unexpected diffusive part may be an indication that there is some liquid-like behavior present already at low temperatures, or that the mean squared displacement is not quite quadratic in time. At the higher temperature, the solution is fully diffusive with approximately log-normally distributed diffusive modes. There is a shoulder on the right side, which the $F$-test has determined should not be smoothed out. These results agree reasonably well with our expectations; the glass-liquid transition can be identified by a shift from mostly ballistic to fully diffusive dynamics, and the positions of the peaks agree with the correlation times identified in the original analysis by \citeauthor{frenzel_2021}\cite{frenzel_2021}.

In figure \ref{fig:ASW_2step}, we show the analysis of our amorphous ice sample. The measurement is taken at 125 K, close to a suspected glass transition temperature~\cite{handa_1988}. The $g_2$ data (figure \ref{fig:ASW_2step}, left panel) shows some complex dynamics with two distinct steps.  We again try a mix of glassy and diffusive dynamics, equation (\ref{eq:ball_diff_g2}), and expect the analysis to show a separation of the fast from the slow part, along with some indication of how well each one fits with the diffusive vs. ballistic character. In the analysis results (top right panel and bottom row of figure \ref{fig:ASW_2step}), the thin spikes in $\Ddist$ and $\vdist$ at fast rates ($D>10$ Å\textsuperscript{2}/s, $v>$1 Å/s) correspond to the first decay-step in $g_2$ and $\isf$. The slow decay is mostly ballistic, while the fast decay comes out as mixed diffusive/ballistic (the diffusive part of $\isf$ has a subtle bump at early times). It should be noted that, in this case, the decomposition of the slow decay is not based on very many data points, and is highly sensitive to the chosen parameter bounds. For example, we do not usually allow negative baselines, but if we relax that restriction, a solution with slow diffusion and negative baselines for the low $q$ values is a better fit than the one presented here. The decomposition of the fast part, however, comes out similar regardless.  

\input{figures_tex/fig_ASW_2step}

\newpage

%% file: figures_tex/fig_PRE2021_2comp.tex
\begin{figure}
    \includegraphics{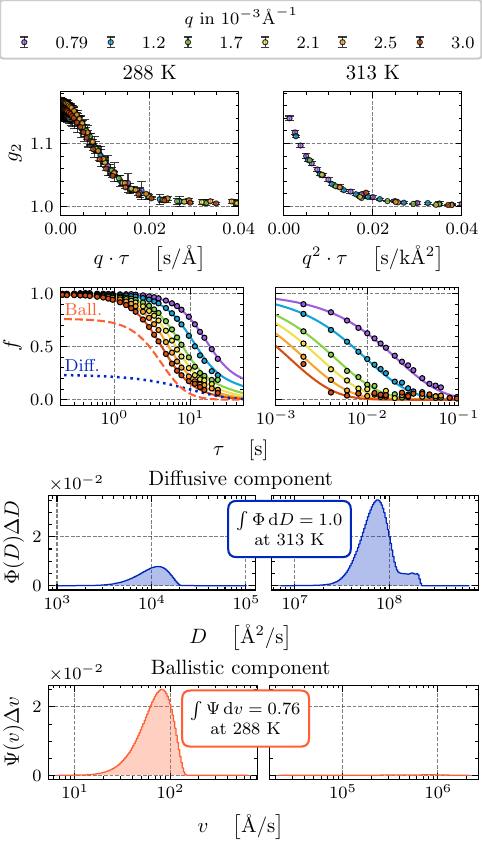}
    \caption{Two-component decomposition, $\smash{f=\int\Ddist\e^{-q^2D\tau}\dd D + \int\vdist\e^{-(qv\tau)^2}}\dd v$, of XPCS data. A dilute suspensions of colloidal particles is measured at two temperatures~\protect\cite{frenzel_2021}. The plots show six select $q$ values out of 12 in the data-set.
    \underline{Left column:} Low temperature measurement.
    \underline{Right column:} High temperature measurement.
    \underline{Top row:} Intensity autocorrelation data plotted against $q$-scaled time axes.
    \underline{Second row:} Best fit intermediate scattering function (solid lines) and data-points corrected by best fit contrast and baseline ($\smash{\sqrt{(g_2-1-b)/\beta}}$). For the low temperature measurement, the ballistic and diffusive components are plotted as separate lines for the highest $q$ ($\smash{3.0\cdot10^{-3}}$ Å\textsuperscript{-1}). 
    \underline{Bottom two rows:} Solution distributions of diffusive and ballistic modes. Rectangles are plotted with height $\Ddist_{m}w_m$ and width $w_m$, such that their areas appear correct on the paper when plotted against a logarithmic $x$-axis.}
    \label{fig:PRE2021_2comp}
\end{figure}

%% file: figures_tex/fig_ASW_2step.tex
\begin{figure}
    \includegraphics{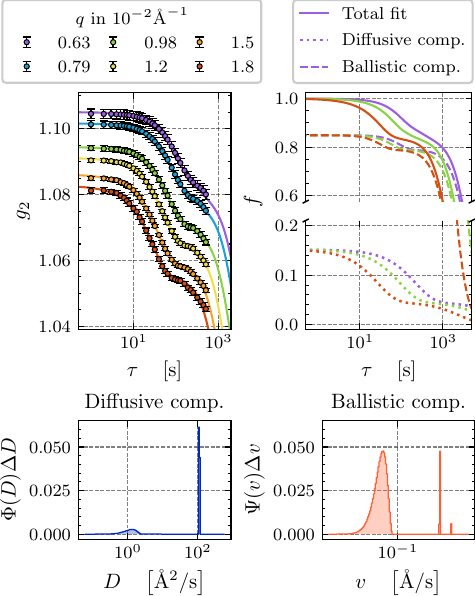}
    \caption{XPCS measurement on amorphous ice decomposed into a diffusive and a ballistic part, $\smash{f=\int\Ddist\e^{-q^2D\tau}\dd D+\int\vdist\e^{-(qv\tau)^2}\dd v}$. Shown are six select $q$ values out of 12 in the data-set (0.0017 to 0.018 Å\textsuperscript{-1}).
    \underline{Top left:} Intensity autocorrelation data and lines of fit.
    \underline{Top right:} The model intermediate scattering function (three select $q$ curves), broken up by component (diffusive/ballistic). \underline{Bottom:} Solution distributions of diffusive and ballistic modes. Rectangles are plotted with height $\Ddist_{m}w_m$ and width $w_m$, such that the areas appear correct on the paper when plotted against a logarithmic $x$-axis.}
    \label{fig:ASW_2step}
\end{figure}

%% file: section_text_files/special_cases.tex
The model fitting described above can be simplified by introducing stronger conditions on the density functions $\phi_l$ in equation (\ref{eq:general_g2}). We previously solved for arbitrary functions (given some boundary conditions and constraints), approximated by a finite number of fitting coefficients. Another option is to substitute these density functions with suitable analytical expressions, in terms of a  small number of distribution parameters. For some particular choices of inverse transform and transform kernel, the intermediate scattering function even has a known closed form. This lets us dispense with the numerical integral altogether. In either case, there is no need for regularization techniques discussed earlier.

\subsection{Analytical expressions for density functions}

Our analysis of the high temperature colloidal particles gave a (nearly) monomodal distribution of diffusivities which, by construction, goes to zero at low and at high rates (see figure~\ref{fig:PRE2021_2comp}). We now seek an analytical expression to stand in for $\Phi(D)$ . We chose the PDF of a log-normal distribution, since
% $\lognorm(s;\mu,\sigma)=1/(2\sigma\sqrt{2\pi})\exp{(-(\ln s-\mu)^2/(2\sigma^2))}$
it exhibits the correct modal and asymptotic behavior, and our model expression \eqref{eq:general_g2} then becomes
\begin{equation}
    \label{eq:simplified_dist_model}
    g_2(q,\tau) = 1 + b(q) + \beta(q)\left(
    \int \frac{1}{s\sigma\sqrt{2\pi}}\e^{-(\ln s-\mu)^2/(2\sigma^2)- q^2 s \tau}\dd s\right)^2
\end{equation}
(we have dropped the ballistic component).
\iffalse
\begin{equation}
    \label{eq:simplified_dist_model}
    {g_2}_{ij} = 1 + b_i + \beta_i\left(
    \int \lognorm(s;\mu_D,\sigma_D^2)\e^{-s D_0 q_i^2 \tau_j}\dd s\right)^2.
\end{equation}
\begin{align}
    \label{eq:simplified_dist_model}
    {g_2}_{ij} = 1 + b_i + \beta_i\bigg(A
        &\int \lognorm(s;\mu_D,\sigma_D^2)
        \e^{-s D_0 q_i^2 \tau_j}\dd s\\
        +\left(1-A\right)&\int\lognorm(s;\mu_v,\sigma_v^2)
        \e^{-s\left(v_0 q_i \tau_j\right)^2}\dd s\bigg)^2.
\end{align}
\fi
This involves $2Q+2$ free parameters to be fitted to data (per $q$ baseline, contrast, and the two distribution parameters $\mu$, $\sigma$). For this specific measurement, we fit 24 parameters to 374 data points.

The integral in equation \eqref{eq:simplified_dist_model} may be evaluated by any method for numerical integration. We found that a practical approach is to fix the number of integrand evaluation points (we use 30 points), then --- on each evaluation of the objective function~--- generate a logarithmically spaced $s$-vector between the 0.0001st and 99.99th percentile and integrate between those bounds using the trapezoid rule. Assigning the integration points dynamically in this manner guarantees that the bulk of the population is accounted for, even as the distribution parameters change during optimization.

The results are contrasted with a CONTIN-style solution in figure~\ref{fig:PRE2021_method_comparison}. Similar to earlier findings~\cite{andrews_2018a}, we see that the more complicated modeling procedure does not necessarily give smaller residuals. This is expected, and is just due to the regularizer doing its intended job. If desired, the residuals of the CONTIN model can be reduced by picking a lower threshold for the criterion \eqref{eq:f-test} (at the cost of a potentially more complex solution). Notably, the log-normal model does not suffer from regularizer-induced biases. In a simple case like this, the simplified approach can provide a plausible decomposition, but it obviously fails when the modality of the distribution is unknown. 

\input{figures_tex/fig_PRE2021_method_comparison}

\subsection{Global fitting of stretched/compressed exponential functions}

 The Kohlrausch–Williams–Watts (or stretched exponential) function,
 \begin{equation}
    \label{eq:stretch}
    \e^{-\left(\Gamma\tau\right)^\gamma},\quad 0<\gamma< 1,
\end{equation}
is commonly used for fitting relaxation rates in correlation data. The inverse Laplace transform of the stretched exponential is a monomodal density function which we will refer to as Pollard's density function $\pollard(x;\gamma)$. It can be computed from the integral expression \cite{pollard_1946}
\begin{equation}
    \label{eq:pollard}
    \pollard(x;\gamma) = \frac{1}{\pi}\int_0^\infty
    \e^{-xy}\e^{-y^\gamma\cos \pi\gamma}\sin\left(y^\gamma\sin\pi\gamma\right)\dd y.
\end{equation}
An observed stretched exponential can be explained by superimposed relaxations according to 
\begin{equation}
    \e^{-\left(\Gamma\tau\right)^\gamma} = \int_0^\infty \pollard(x;\gamma)\e^{-x \Gamma\tau}\dd x,\qquad0<\gamma<1.
\end{equation}
For $\gamma$ close to 1, the Pollard density gets increasingly sharp approaching a $\delta$-function centered on $x=1$. The compressed case, where $\gamma>1$, can similarly be expressed as a linear combination of Gaussian decays,\cite{hansen_2013}
\begin{equation}
    \e^{-\left(\Gamma\tau\right)^\gamma} = \int_0^\infty \pollard(x;\gamma)\e^{- x\left(\Gamma\tau\right)^2}\dd x,
    \qquad1<\gamma<2.
\end{equation}
Using this analytical solution, information about underlying distributions of decays is inferred from a simple two-parameter curve-fit without resorting to numerical solution of integral equations.

Traditionally, in DLS as well as in XPCS, the stretched exponential (\ref{eq:stretch}) is optimized against $g_2$ data for each $q$-bin by curve fitting, and then the relaxation rates $\Gamma$ are studied as a function of $q$. A characteristic diffusion rate, $\overline{D}$, can be extracted by a subsequent fit according to $\Gamma = \overline{D} q^2$. Here, we will instead take a global approach. By fitting the model
\begin{equation}
    \label{eq:global_KWW}
    % g_2(q,\tau) = 1+b(q)+\beta(q)\e^{-2\left(\Gamma q^\epsilon\tau \right)^\gamma}
    g_2(q,\tau) = 1+b(q)+\beta(q)\e^{-2\left( q^2 \overline{D} \tau \right)^\gamma}
\end{equation}
to our data in a surface fit manner ($q$ and $\tau$ are the independent variables, $\overline{D}$ and $\gamma$ are fit parameters), we can solve for the best fit Pollard-distributed population of diffusion rates directly. In this case, the underlying expression for the dynamic scattering function would be
\begin{equation}
    \isf(q,\tau)
    % =\e^{-(\Gamma q^\epsilon\tau)^\gamma}
    %     =\frac{\eta s^{\eta-1}}{\Gamma^\eta}
    % \int\pollard\left(s^\eta/\Gamma^\eta;\gamma\right)
    % \e^{-q^{\epsilon\eta}(s\tau)^\eta}\dd s
    % \quad \text{with}\quad \eta=\lceil\gamma\rceil,\,0<\gamma\leq2.
    =\frac{1}{\overline{D}}
    \int\pollard\left(s/\overline{D};\gamma\right)
    \e^{-q^2 s\tau} \dd s. %,\quad0<\gamma\leq 1.
\end{equation}

Figure~\ref{fig:PRE2021_method_comparison} includes results from the global KWW analysis applied to the high temperature colloids from figure \ref{fig:PRE2021_2comp}. While the interpretation in terms of diffusive modes is quite different from the log-normal model, the distribution of the residuals is nearly identical (orange vs. green curves/points). Dynamic scattering generated by an underlying Pollard vs. a log-normal density hence appear phenomenologically indistinct, and preference for one or the other will depend on the specific goals of the analysis. The advantage of the KWW model lies in its exceptionally convenient time domain representation, while the Laplace domain expression \eqref{eq:pollard} is cumbersome. The log-normal model does not transform into time as cleanly, but has a simpler form in the Laplace domain.

The global KWW technique can be adapted for more complex data, with several suspected stretched or compressed dynamic components. For example, with the fit 
\begin{equation}
    g_2 = 1+b(q)+\beta(q)\left(
    A\e^{-\left(q\overline{v}\tau\right)^{\gamma_1}}
    +\left(1-A\right)\e^{-\left(q^2\overline{D}\tau \right)^{\gamma_2}}\right)^2,\quad0<\gamma_1\leq1,\quad1<\gamma_2\leq2,
\end{equation}
$g_2$ is decomposed into a ballistic (compressed) and a diffusive (stretched) component with characteristic velocities/diffusion rates $v$, $D$. For an example of this multi-component global fitting technique, see the recent paper by \citeauthor{karina_2025}\cite{karina_2025} 

%% file: figures_tex/fig_PRE2021_method_comparison.tex
\begin{figure}
    \includegraphics{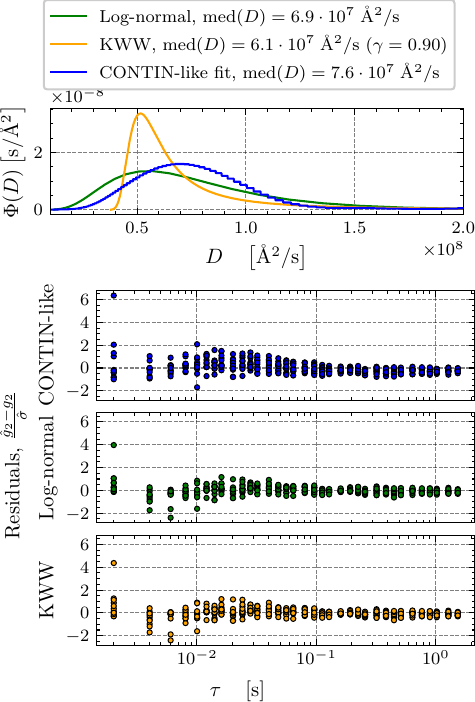}
    \caption{Comparison of the three methods discussed in this paper. A diffusive model, $\smash{f=\int\Ddist\e^{-Dq^2\tau}\dd D}$, is fitted to the high temperature data from figure \ref{fig:PRE2021_2comp}. The CONTIN-like fit (blue) solves for an arbitrary density function $\Phi(D)$, discretized into fitting coefficients. The log-normal fit (green) fits a log-normal distribution, equation \eqref{eq:simplified_dist_model}. The KWW-fit (orange) is a global stretched exponential fit, equation \eqref{eq:global_KWW}. The median, $\mathrm{med}(D)$, of each distribution is printed in the legend.}
    \label{fig:PRE2021_method_comparison}
\end{figure}

%% file: section_text_files/conclusions.tex
We have presented a new framework for analyzing intensity autocorrelation data using global, $q$-dependent inverse transforms. By directly modeling the nonlinear $g_2$ function and implementing a flexible, regularized fitting strategy, our approach enables robust decomposition of complex dynamics into, e.g., diffusive and ballistic components. Our software builds on and extends prior efforts, such as CONTIN and MULTIQ, while introducing modern optimization techniques and tools for selecting the regularization parameter. Example applications to colloidal suspensions and amorphous ice demonstrate the method’s ability to resolve multi-component dynamical behavior and provide insight into underlying relaxation processes. Our implementation, including full source code and example data, is openly available to facilitate further development and adoption in the community.\cite{g2_decomposer_gh} We anticipate that this tool will support a broad range of studies in soft and disordered matter, advancing the quantitative analysis of dynamics probed by XPCS and related techniques.

%% file: section_text_files/data_availability.tex
The Matlab functions and example analysis scripts described here are published together with the example data from P10, PETRA III, as a \texttt{git} repository, publicly available on GitHub~\cite{g2_decomposer_gh}. The code is tested with Matlab and Optimization Toolbox ver. 2024b. This text concerns the version 1.0.1 of our software. We employ an open-ended license; anyone is free to use, modify and re-distribute the software package.

%% file: section_text_files/acknowledgement.tex
We acknowledge Deutsches Elektronen-Synchrotron DESY (Hamburg, Germany), a member of the Helmholtz Association HGF, for the provision of experimental facilities. Parts of this research were carried out at PETRA III, beamline P10. Beamtime was allocated for proposals I-20180073 and T-20220749 This research was supported in part through the Maxwell computational resources operated at DESY, Hamburg, Germany. T.E. acknowledges funding by the Centre for Molecular Water Science (CMWS) within an Early Science Project and by the Deutsche Forschungsgemeinschaft (DFG, German Research Foundation) - SFB1552 - 465145163, project Q3. C.M.T. acknowledges funding by the Centre for Molecular Water Science (CMWS) within an Early Science Project, by PIER - Partnership of University Hamburg and DESY within a PIER Seed Project (PIF-2024-05) and by the Austrian Academy of Sciences (ÖAW) and the Austrian Science Fund (FWF) within the Disruptive Innovation-Early Career Seed Money Grant. We also acknowledge the scientific exchange and support of the CMWS. We thank Werner Steffen and Fabian Westermeier for scientific discussions.

%% file: main.bbl
\providecommand{\latin}[1]{#1}
\makeatletter
\providecommand{\doi}
  {\begingroup\let\do\@makeother\dospecials
  \catcode`\{=1 \catcode`\}=2 \doi@aux}
\providecommand{\doi@aux}[1]{\endgroup\texttt{#1}}
\makeatother
\providecommand*\mcitethebibliography{\thebibliography}
\csname @ifundefined\endcsname{endmcitethebibliography}
  {\let\endmcitethebibliography\endthebibliography}{}
\begin{mcitethebibliography}{34}
\providecommand*\natexlab[1]{#1}
\providecommand*\mciteSetBstSublistMode[1]{}
\providecommand*\mciteSetBstMaxWidthForm[2]{}
\providecommand*\mciteBstWouldAddEndPuncttrue
  {\def\EndOfBibitem{\unskip.}}
\providecommand*\mciteBstWouldAddEndPunctfalse
  {\let\EndOfBibitem\relax}
\providecommand*\mciteSetBstMidEndSepPunct[3]{}
\providecommand*\mciteSetBstSublistLabelBeginEnd[3]{}
\providecommand*\EndOfBibitem{}
\mciteSetBstSublistMode{f}
\mciteSetBstMaxWidthForm{subitem}{(\alph{mcitesubitemcount})}
\mciteSetBstSublistLabelBeginEnd
  {\mcitemaxwidthsubitemform\space}
  {\relax}
  {\relax}

\bibitem[Berne and Pecora(2000)Berne, and Pecora]{berne_dls}
Berne,~B.; Pecora,~R. \emph{Dynamic Light Scattering: With Applications to
  Chemistry, Biology, and Physics}; Dover Books on Physics Series; Dover
  Publications, 2000\relax
\mciteBstWouldAddEndPuncttrue
\mciteSetBstMidEndSepPunct{\mcitedefaultmidpunct}
{\mcitedefaultendpunct}{\mcitedefaultseppunct}\relax
\EndOfBibitem
\bibitem[Lehmkühler \latin{et~al.}(2021)Lehmkühler, Roseker, and
  Grübel]{lehmkuehler_2021}
Lehmkühler,~F.; Roseker,~W.; Grübel,~G. From Femtoseconds to
  Hours—Measuring Dynamics over 18 Orders of Magnitude with Coherent X-rays.
  \emph{Applied Sciences} \textbf{2021}, \emph{11}\relax
\mciteBstWouldAddEndPuncttrue
\mciteSetBstMidEndSepPunct{\mcitedefaultmidpunct}
{\mcitedefaultendpunct}{\mcitedefaultseppunct}\relax
\EndOfBibitem
\bibitem[Sandy \latin{et~al.}(2018)Sandy, Zhang, and Lurio]{sandy_2018}
Sandy,~A.~R.; Zhang,~Q.; Lurio,~L.~B. Hard X-Ray Photon Correlation
  Spectroscopy Methods for Materials Studies. \emph{Annual Review of Materials
  Research} \textbf{2018}, \emph{48}, 167--190\relax
\mciteBstWouldAddEndPuncttrue
\mciteSetBstMidEndSepPunct{\mcitedefaultmidpunct}
{\mcitedefaultendpunct}{\mcitedefaultseppunct}\relax
\EndOfBibitem
\bibitem[Madsen \latin{et~al.}(2020)Madsen, Fluerasu, and Ruta]{madsen_2018}
Madsen,~A.; Fluerasu,~A.; Ruta,~B. In \emph{Synchrotron Light Sources and
  Free-Electron Lasers: Accelerator Physics, Instrumentation and Science
  Applications}; Jaeschke,~E.~J., Khan,~S., Schneider,~J.~R., Hastings,~J.~B.,
  Eds.; Springer International Publishing: Cham, 2020; pp 1989--2018\relax
\mciteBstWouldAddEndPuncttrue
\mciteSetBstMidEndSepPunct{\mcitedefaultmidpunct}
{\mcitedefaultendpunct}{\mcitedefaultseppunct}\relax
\EndOfBibitem
\bibitem[Shpyrko(2014)]{shpyrko_2014}
Shpyrko,~O.~G. {X-ray photon correlation spectroscopy}. \emph{Journal of
  Synchrotron Radiation} \textbf{2014}, \emph{21}, 1057--1064\relax
\mciteBstWouldAddEndPuncttrue
\mciteSetBstMidEndSepPunct{\mcitedefaultmidpunct}
{\mcitedefaultendpunct}{\mcitedefaultseppunct}\relax
\EndOfBibitem
\bibitem[Sinha \latin{et~al.}(2014)Sinha, Jiang, and Lurio]{sinha_2014}
Sinha,~S.~K.; Jiang,~Z.; Lurio,~L.~B. X-ray Photon Correlation Spectroscopy
  Studies of Surfaces and Thin Films. \emph{Advanced Materials} \textbf{2014},
  \emph{26}, 7764--7785\relax
\mciteBstWouldAddEndPuncttrue
\mciteSetBstMidEndSepPunct{\mcitedefaultmidpunct}
{\mcitedefaultendpunct}{\mcitedefaultseppunct}\relax
\EndOfBibitem
\bibitem[Perakis and Gutt(2020)Perakis, and Gutt]{perakis_2020}
Perakis,~F.; Gutt,~C. Towards molecular movies with X-ray photon correlation
  spectroscopy. \emph{Phys. Chem. Chem. Phys.} \textbf{2020}, \emph{22},
  19443--19453\relax
\mciteBstWouldAddEndPuncttrue
\mciteSetBstMidEndSepPunct{\mcitedefaultmidpunct}
{\mcitedefaultendpunct}{\mcitedefaultseppunct}\relax
\EndOfBibitem
\bibitem[Grübel and Zontone(2004)Grübel, and Zontone]{grubel_2004}
Grübel,~G.; Zontone,~F. Correlation spectroscopy with coherent X-rays.
  \emph{Journal of Alloys and Compounds} \textbf{2004}, \emph{362}, 3--11,
  Proceedings of the Sixth International School and Symposium on Synchrotron
  Radiation in Natural Science (ISSRNS)\relax
\mciteBstWouldAddEndPuncttrue
\mciteSetBstMidEndSepPunct{\mcitedefaultmidpunct}
{\mcitedefaultendpunct}{\mcitedefaultseppunct}\relax
\EndOfBibitem
\bibitem[Koppel(1972)]{koppel_1972}
Koppel,~D.~E. Analysis of Macromolecular Polydispersity in Intensity
  Correlation Spectroscopy: The Method of Cumulants. \emph{The Journal of
  Chemical Physics} \textbf{1972}, \emph{57}, 4814--4820\relax
\mciteBstWouldAddEndPuncttrue
\mciteSetBstMidEndSepPunct{\mcitedefaultmidpunct}
{\mcitedefaultendpunct}{\mcitedefaultseppunct}\relax
\EndOfBibitem
\bibitem[Provencher(1982)]{provencher_1982a}
Provencher,~S.~W. A constrained regularization method for inverting data
  represented by linear algebraic or integral equations. \emph{Computer Physics
  Communications} \textbf{1982}, \emph{27}, 213--227\relax
\mciteBstWouldAddEndPuncttrue
\mciteSetBstMidEndSepPunct{\mcitedefaultmidpunct}
{\mcitedefaultendpunct}{\mcitedefaultseppunct}\relax
\EndOfBibitem
\bibitem[Provencher(1979)]{provencher_1979}
Provencher,~S.~W. Inverse problems in polymer characterization: Direct analysis
  of polydispersity with photon correlation spectroscopy. \emph{Die
  Makromolekulare Chemie} \textbf{1979}, \emph{180}, 201--209\relax
\mciteBstWouldAddEndPuncttrue
\mciteSetBstMidEndSepPunct{\mcitedefaultmidpunct}
{\mcitedefaultendpunct}{\mcitedefaultseppunct}\relax
\EndOfBibitem
\bibitem[Provencher and Štêpánek(1996)Provencher, and
  Štêpánek]{provencher_1996}
Provencher,~S.~W.; Štêpánek,~P. Global Analysis of Dynamic Light Scattering
  Autocorrelation Functions. \emph{Particle \& Particle Systems
  Characterization} \textbf{1996}, \emph{13}, 291--294\relax
\mciteBstWouldAddEndPuncttrue
\mciteSetBstMidEndSepPunct{\mcitedefaultmidpunct}
{\mcitedefaultendpunct}{\mcitedefaultseppunct}\relax
\EndOfBibitem
\bibitem[Andrews \latin{et~al.}(2018)Andrews, Narayanan, Zhang, Kuzmenko, and
  Ilavsky]{andrews_2018a}
Andrews,~R.~N.; Narayanan,~S.; Zhang,~F.; Kuzmenko,~I.; Ilavsky,~J. {Inverse
  transformation: unleashing spatially heterogeneous dynamics with an
  alternative approach to XPCS data analysis}. \emph{Journal of Applied
  Crystallography} \textbf{2018}, \emph{51}, 35--46\relax
\mciteBstWouldAddEndPuncttrue
\mciteSetBstMidEndSepPunct{\mcitedefaultmidpunct}
{\mcitedefaultendpunct}{\mcitedefaultseppunct}\relax
\EndOfBibitem
\bibitem[Marino(2007)]{marino_2007_mlfe}
Marino,~I.-G. rilt.
  \url{https://www.mathworks.com/matlabcentral/fileexchange/6523-rilt},
  2007\relax
\mciteBstWouldAddEndPuncttrue
\mciteSetBstMidEndSepPunct{\mcitedefaultmidpunct}
{\mcitedefaultendpunct}{\mcitedefaultseppunct}\relax
\EndOfBibitem
\bibitem[Liénard \latin{et~al.}(2022)Liénard, Freyssingeas, and
  Borgnat]{lienard_2022}
Liénard,~F.; Freyssingeas,~{\'E}.; Borgnat,~P. A multiscale time-Laplace
  method to extract relaxation times from non-stationary dynamic light
  scattering signals. \emph{The Journal of Chemical Physics} \textbf{2022},
  \emph{156}, 224901\relax
\mciteBstWouldAddEndPuncttrue
\mciteSetBstMidEndSepPunct{\mcitedefaultmidpunct}
{\mcitedefaultendpunct}{\mcitedefaultseppunct}\relax
\EndOfBibitem
\bibitem[Liénard \latin{et~al.}(2022)Liénard, Freyssingeas, and
  Borgnat]{lienard_2022_mlfe}
Liénard,~F.; Freyssingeas,~E.; Borgnat,~P. {RILT} fmincon.
  \url{https://de.mathworks.com/matlabcentral/fileexchange/119708-rilt-fmincon},
  2022\relax
\mciteBstWouldAddEndPuncttrue
\mciteSetBstMidEndSepPunct{\mcitedefaultmidpunct}
{\mcitedefaultendpunct}{\mcitedefaultseppunct}\relax
\EndOfBibitem
\bibitem[Voigt and Hess(1994)Voigt, and Hess]{voigt_1994}
Voigt,~H.; Hess,~S. Comparison of the intensity correlation function and the
  intermediate scattering function of fluids: a molecular dynamics study of the
  Siegert relation. \emph{Physica A: Statistical Mechanics and its
  Applications} \textbf{1994}, \emph{202}, 145--164\relax
\mciteBstWouldAddEndPuncttrue
\mciteSetBstMidEndSepPunct{\mcitedefaultmidpunct}
{\mcitedefaultendpunct}{\mcitedefaultseppunct}\relax
\EndOfBibitem
\bibitem[Provencher \latin{et~al.}(1978)Provencher, Hendrix, De~Maeyer, and
  Paulussen]{provencher_1978}
Provencher,~S.~W.; Hendrix,~J.; De~Maeyer,~L.; Paulussen,~N. Direct
  determination of molecular weight distributions of polystyrene in cyclohexane
  with photon correlation spectroscopy. \emph{The Journal of Chemical Physics}
  \textbf{1978}, \emph{69}, 4273--4276\relax
\mciteBstWouldAddEndPuncttrue
\mciteSetBstMidEndSepPunct{\mcitedefaultmidpunct}
{\mcitedefaultendpunct}{\mcitedefaultseppunct}\relax
\EndOfBibitem
\bibitem[Ladd-Parada \latin{et~al.}(2022)Ladd-Parada, Li, Karina, Kim, Perakis,
  Reiser, Dallari, Striker, Sprung, Westermeier, Grübel, Nilsson, Lehmkühler,
  and Amann-Winkel]{ladd_parada_2022}
Ladd-Parada,~M.; Li,~H.; Karina,~A.; Kim,~K.~H.; Perakis,~F.; Reiser,~M.;
  Dallari,~F.; Striker,~N.; Sprung,~M.; Westermeier,~F.; Grübel,~G.;
  Nilsson,~A.; Lehmkühler,~F.; Amann-Winkel,~K. Using coherent X-rays to
  follow dynamics in amorphous ices. \emph{Environ. Sci.: Atmos.}
  \textbf{2022}, \emph{2}, 1314--1323\relax
\mciteBstWouldAddEndPuncttrue
\mciteSetBstMidEndSepPunct{\mcitedefaultmidpunct}
{\mcitedefaultendpunct}{\mcitedefaultseppunct}\relax
\EndOfBibitem
\bibitem[Weese(1993)]{weese_1993}
Weese,~J. A regularization method for nonlinear ill-posed problems.
  \emph{Computer Physics Communications} \textbf{1993}, \emph{77},
  429--440\relax
\mciteBstWouldAddEndPuncttrue
\mciteSetBstMidEndSepPunct{\mcitedefaultmidpunct}
{\mcitedefaultendpunct}{\mcitedefaultseppunct}\relax
\EndOfBibitem
\bibitem[Ye(1998)]{ye_1998}
Ye,~J. On Measuring and Correcting the Effects of Data Mining and Model
  Selection. \emph{J. Am. Stat. Assoc.} \textbf{1998}, \emph{93},
  120--131\relax
\mciteBstWouldAddEndPuncttrue
\mciteSetBstMidEndSepPunct{\mcitedefaultmidpunct}
{\mcitedefaultendpunct}{\mcitedefaultseppunct}\relax
\EndOfBibitem
\bibitem[Andrews \latin{et~al.}(2018)Andrews, Narayanan, Zhang, Kuzmenko, and
  Ilavsky]{andrews_2018b}
Andrews,~R.~N.; Narayanan,~S.; Zhang,~F.; Kuzmenko,~I.; Ilavsky,~J. {{\it
  CONTIN XPCS}: software for inverse transform analysis of X-ray photon
  correlation spectroscopy dynamics}. \emph{Journal of Applied Crystallography}
  \textbf{2018}, \emph{51}, 205--209\relax
\mciteBstWouldAddEndPuncttrue
\mciteSetBstMidEndSepPunct{\mcitedefaultmidpunct}
{\mcitedefaultendpunct}{\mcitedefaultseppunct}\relax
\EndOfBibitem
\bibitem[Nocedal and Wright(2006)Nocedal, and Wright]{nocedal_2006}
Nocedal,~J.; Wright,~S. \emph{Numerical Optimization}, 2nd ed.; Springer
  Science+Business Media, 2006; Chapter 18\relax
\mciteBstWouldAddEndPuncttrue
\mciteSetBstMidEndSepPunct{\mcitedefaultmidpunct}
{\mcitedefaultendpunct}{\mcitedefaultseppunct}\relax
\EndOfBibitem
\bibitem[Provencher(1982)]{provencher_1982b}
Provencher,~S.~W. CONTIN: A general purpose constrained regularization program
  for inverting noisy linear algebraic and integral equations. \emph{Computer
  Physics Communications} \textbf{1982}, \emph{27}, 229--242\relax
\mciteBstWouldAddEndPuncttrue
\mciteSetBstMidEndSepPunct{\mcitedefaultmidpunct}
{\mcitedefaultendpunct}{\mcitedefaultseppunct}\relax
\EndOfBibitem
\bibitem[Frenzel \latin{et~al.}(2021)Frenzel, Dartsch, Balaguer, Westermeier,
  Gr\"ubel, and Lehmk\"uhler]{frenzel_2021}
Frenzel,~L.; Dartsch,~M.; Balaguer,~G.~M.; Westermeier,~F.; Gr\"ubel,~G.;
  Lehmk\"uhler,~F. Glass-liquid and glass-gel transitions of soft-shell
  particles. \emph{Phys. Rev. E} \textbf{2021}, \emph{104}, L012602\relax
\mciteBstWouldAddEndPuncttrue
\mciteSetBstMidEndSepPunct{\mcitedefaultmidpunct}
{\mcitedefaultendpunct}{\mcitedefaultseppunct}\relax
\EndOfBibitem
\bibitem[Burton and Oliver(1935)Burton, and Oliver]{burton_1935}
Burton,~E.~F.; Oliver,~W.~F. X-Ray Diffraction Patterns of Ice. \emph{Nature}
  \textbf{1935}, \emph{135}, 505–506\relax
\mciteBstWouldAddEndPuncttrue
\mciteSetBstMidEndSepPunct{\mcitedefaultmidpunct}
{\mcitedefaultendpunct}{\mcitedefaultseppunct}\relax
\EndOfBibitem
\bibitem[Mayer and Pletzer(1986)Mayer, and Pletzer]{mayer_1986}
Mayer,~E.; Pletzer,~R. Astrophysical implications of amorphous ice—a
  microporous solid. \emph{Nature} \textbf{1986}, \emph{319}, 298–301\relax
\mciteBstWouldAddEndPuncttrue
\mciteSetBstMidEndSepPunct{\mcitedefaultmidpunct}
{\mcitedefaultendpunct}{\mcitedefaultseppunct}\relax
\EndOfBibitem
\bibitem[Li \latin{et~al.}(2021)Li, Karina, Ladd-Parada, Sp\"{a}h, Perakis,
  Benmore, and Amann-Winkel]{li_2021}
Li,~H.; Karina,~A.; Ladd-Parada,~M.; Sp\"{a}h,~A.; Perakis,~F.; Benmore,~C.;
  Amann-Winkel,~K. Long-Range Structures of Amorphous Solid Water. \emph{The
  Journal of Physical Chemistry B} \textbf{2021}, \emph{125},
  13320–13328\relax
\mciteBstWouldAddEndPuncttrue
\mciteSetBstMidEndSepPunct{\mcitedefaultmidpunct}
{\mcitedefaultendpunct}{\mcitedefaultseppunct}\relax
\EndOfBibitem
\bibitem[Eklund \latin{et~al.}(2025)Eklund, Tonauer, Lehmkühler, and
  Amann-Winkel]{g2_decomposer_gh}
Eklund,~T.; Tonauer,~C.~M.; Lehmkühler,~F.; Amann-Winkel,~K. {$g_2$
  decomposer}. 2025; \url{https://github.com/elkund/g2_decomposer}\relax
\mciteBstWouldAddEndPuncttrue
\mciteSetBstMidEndSepPunct{\mcitedefaultmidpunct}
{\mcitedefaultendpunct}{\mcitedefaultseppunct}\relax
\EndOfBibitem
\bibitem[Handa and Klug(1988)Handa, and Klug]{handa_1988}
Handa,~Y.~P.; Klug,~D.~D. Heat capacity and glass transition behavior of
  amorphous ice. \emph{J. Phys. Chem. A} \textbf{1988}, \emph{92},
  3323--3325\relax
\mciteBstWouldAddEndPuncttrue
\mciteSetBstMidEndSepPunct{\mcitedefaultmidpunct}
{\mcitedefaultendpunct}{\mcitedefaultseppunct}\relax
\EndOfBibitem
\bibitem[Pollard(1946)]{pollard_1946}
Pollard,~H. The representation of $e^{-x^{\lambda}}$ as a Laplace integral.
  \emph{Bulletin of the American Mathematical Society} \textbf{1946},
  \emph{52}, 908–910\relax
\mciteBstWouldAddEndPuncttrue
\mciteSetBstMidEndSepPunct{\mcitedefaultmidpunct}
{\mcitedefaultendpunct}{\mcitedefaultseppunct}\relax
\EndOfBibitem
\bibitem[Hansen \latin{et~al.}(2013)Hansen, Gong, and Chen]{hansen_2013}
Hansen,~E.~W.; Gong,~X.; Chen,~Q. Compressed Exponential Response Function
  Arising From a Continuous Distribution of Gaussian Decays – Distribution
  Characteristics. \emph{Macromolecular Chemistry and Physics} \textbf{2013},
  \emph{214}, 844--852\relax
\mciteBstWouldAddEndPuncttrue
\mciteSetBstMidEndSepPunct{\mcitedefaultmidpunct}
{\mcitedefaultendpunct}{\mcitedefaultseppunct}\relax
\EndOfBibitem
\bibitem[Karina \latin{et~al.}(2025)Karina, Li, Eklund, Ladd-Parada, Massani,
  Filianina, Kondedan, Rydh, Holl, Trevorah, Huotari, Bauer, Goy, Striker,
  Dallari, Westermeier, Sprung, Lehmk\"{u}hler, and Amann-Winkel]{karina_2025}
Karina,~A. \latin{et~al.}  Multicomponent dynamics in amorphous ice studied
  using X-ray photon correlation spectroscopy at elevated pressure and
  cryogenic temperatures. \emph{Communications Chemistry} \textbf{2025},
  \emph{8}\relax
\mciteBstWouldAddEndPuncttrue
\mciteSetBstMidEndSepPunct{\mcitedefaultmidpunct}
{\mcitedefaultendpunct}{\mcitedefaultseppunct}\relax
\EndOfBibitem
\end{mcitethebibliography}
